\documentclass[letterpaper, 10 pt, conference]{ieeeconf} 
\usepackage{amsmath} 
\usepackage{amssymb}  
\usepackage{longtable}
\usepackage{tabularx}
\usepackage{array}
\usepackage{booktabs}
\usepackage{textcomp}

\usepackage{graphicx}
\usepackage{booktabs}
\usepackage{cite}
\usepackage{float}
\newcommand*{\QEDA}{\hfill\ensuremath{\blacksquare}}

\usepackage[linesnumbered,boxed,ruled,commentsnumbered]{algorithm2e}
\SetKwRepeat{Do}{do}{while}%
\usepackage{color}

\IEEEoverridecommandlockouts                              

\makeatletter
\def\endthebibliography{%
  \def\@noitemerr{\@latex@warning{Empty `thebibliography' environment}}%
  \endlist
}
\makeatother

\title{\LARGE \bf
Predictive Lane-Change and Routing Coordination in  \vspace{2pt} \\ Bus-Priority Mixed Traffic Corridors

}

\author{Tanlu Liang, Ting Bai, and Andreas A. Malikopoulos~\IEEEmembership{Senior Member, IEEE}
\thanks{This research was supported in part by NSF under Grants CNS-2401007, CMMI-2219761, IIS-2415478, and in part by MathWorks.}\vspace{1.5pt}
\thanks{T. Liang, T. Bai, and A. A. Malikopoulos are with the Information and Decision Science Lab, School of Civil $\&$ Environmental Engineering, Cornell University, Ithaca, New York, USA. E-mails: \{{\tt\small tl933, tingbai, amaliko\}@cornell.edu}}
}
\begin{document}

\maketitle
\thispagestyle{empty}
\pagestyle{empty}

\begin{abstract}
In this paper, we investigate the coordination of vehicle maneuvers in mixed-traffic corridors where connected and automated vehicles, human-driven vehicles, and buses interact under dedicated bus lane operations. We develop a segment-based network coordination framework that jointly optimizes lane-change and routing decisions of connected and automated vehicles to improve dedicated lane utilization while preserving bus priority. 
The proposed framework incorporates a predictive bus-protection mechanism that restricts vehicle access to protected lane segments within a monitoring horizon, together with a utility-driven lane-change strategy that accounts for anticipated travel time gains, downstream routing feasibility, and lane-change stability. By explicitly coupling network-level routing decisions with lane-level interaction control, the method proactively mitigates conflicts on dedicated lanes before congestion effects materialize. The proposed approach is evaluated through microscopic traffic simulations in SUMO using a realistic urban corridor. Simulation results demonstrate that the framework enhances bus schedule adherence and reduces average travel times for both automated and human-driven vehicles, while maintaining stable lane-change behavior without increasing maneuver frequency.
\end{abstract}

\section{Introduction}
The deployment of connected and automated vehicles (CAVs) has experienced significant growth in recent years and is widely recognized as a promising strategy for enhancing transportation system efficiency  \cite{Malikopoulos2020,Malikopoulos2019CDC,Bang2022combined2,10209062}. Their vehicle-to-vehicle (V2V) and vehicle-to-infrastructure (V2I) communications enable more coordinated and responsive traffic operations, thereby creating opportunities to reduce overall travel time and improve road safety \cite{zhao2019enhanced,YI2024100126,le2024distributed2,9683080}. During the prolonged transitional period in which CAVs coexist with human-driven vehicles (HDVs) before fully dominating the transportation system, it is important to understand how to effectively leverage the advantages of CAVs, such as shorter headways, reduced reaction times, and controllable routing, within a heterogeneous traffic environment \cite{10955737}.

Prior studies have demonstrated that implementing dedicated lanes (DL) for CAVs provides an effective way to fully exploit their operational advantages and improve overall traffic efficiency, whereas operating CAVs in mixed lanes with HVs diminishes their potential benefits due to speed variability and disturbance propagation~\cite{11304245}. It was reported that traffic crashes significantly decrease when CAV DL are deployed under high market penetration rates~\cite{SHA2024107424}. Jiang and Shang proposed a dynamic CAV DL allocation method integrated with signal timing and trajectory optimization, showing substantial improvements in traffic efficiency \cite{9774973}. However, constructing new CAV DLs involves high infrastructure costs, and converting existing lanes to CAV exclusive use reduce overall roadway capacity~\cite{systems13110958}. In addition, public acceptance and understanding of CAV lane policies may require a considerable adjustment period.

Dedicated bus lanes (DBLs) have been established in dense urban areas, and various multiplexing strategies have been proposed to further improve its utilization \cite{app131810098}, \cite{https://doi.org/10.1155/2022/4799497}, \cite{su17051991}. For instance, Chen proposed a joint DL strategy that permits CAV access to DBL under controlled conditions, thereby improving DBL utilization and partially separating CAVs from the mixed traffic stream to better exploit their operational advantages \cite{CHEN2020102795}. A review has summarized existing studies on joint DL operations for CAV–bus systems~\cite{11304245}. Xiu et al. proposed a CAV dynamic control strategy in which each CAV monitors nearby buses within a sensing radius and determines whether to enter or vacate the joint DL to avoid bus interference~\cite{https://doi.org/10.1155/2021/2470738}. Liu et al. developed a cellular automata–based mixed-traffic model with mandatory lane-changing rules, enabling intermittent CAV use of the bus lane and systematically evaluating the impacts of CAV penetration rate, bus departure interval, and bus stop density~\cite{9798494}. From a more advanced control perspective, a dynamic multi-function lane management framework was proposed that first allocates bus-priority right-of-way and then applies Monte Carlo tree search (MCTS) based lane access decisions together with trajectory optimization for CAVs~\cite{su16188078}. Chen et al. formulated a cyclic space–time network and sequential optimization approach to optimize AV lane usage and flow allocation along a bus corridor, thereby reducing total travel time \cite{CHEN2020102795}.

At a broader scale, some studies investigate network-level optimization of joint DL facilities \cite{su14148686}, \cite{YangJie}. Existing studies largely fall into (i) local lane access and vehicle-level control, and (ii) static network design and schedule coordination. A unified framework that operates at the network level while proactively coordinating CAV routing decisions with bus priority preservation remains largely underexplored. A prior work partially addressed this challenge by proposing a network level dynamic routing strategy that proactively reroutes CAVs in the presence of joint DL operations~\cite{liang2025coordinatedroutingapproachenhancing}. While effective in mitigating potential bus interference from a macroscopic routing perspective, that framework does not explicitly capture lane-level maneuvering behavior, which may further enhance operational flexibility and system performance.

In this paper, we develop a segment based network coordination framework for mixed traffic environments in which buses and CAVs share the DL. The proposed approach explicitly models and regulates the lane-change behavior of CAVs in the vicinity of bus operations, while routing decisions are coordinated to mitigate potential delays. The key idea is to couple network-level CAV routing with lane-level interaction management so that potential conflicts on DL segments can be mitigated before they materialize. By leveraging real time traffic observations and short horizon flow estimation, the framework dynamically identifies the most suitable CAVs for lane-change based on a comprehensive evaluation of expected travel time save, routing constraints, and the cumulative cost of frequent lane changes. In addition, a bus-centered protection mechanism proactively regulates CAV access to DL segments to prevent bus interference and preserve schedule adherence. The main contributions are summarized as twofold. (i) We formulate a lane-level control problem for CAVs operating in a mixed traffic network with joint DL usage. (ii) We develop a prediction-driven monitoring and control mechanism that enables timely lane-change regulation together with supportive CAV rerouting, allowing protection of DL operations under dynamic traffic conditions. Simulation experiments conducted in the SUMO environment using a realistic San Francisco network demonstrate that the proposed framework significantly improves bus operational reliability while maintaining efficient utilization of the DL by CAVs, and the overall number of lane changes remains within a reasonable range without introducing additional safety risks.

\section{Problem Statement}
Consider an urban transportation network represented by a directed graph \(G\!=\!(V, E)\), 
where \(V\) denotes the set of nodes and \(E\!\subseteq\! V\! \times\!V\) 
denotes the set of directed edges. Each edge \((v, v')\!\in\!E\) represents a 
road segment from node \(v\) to node \(v'\) and contains two lanes in the same direction, denoted by a left lane \(e^L(v, v')\) and a right lane \(e^R(v, v')\). 
The left lane is always a general-purpose lane (GPL), while the right lane may serve either as a GPL or as a DL. We capture this infrastructure property with a binary variable
\begin{equation}
\delta(v, v') =
\begin{cases}
1, & \text{if } e^R(v, v') \text{ is a DL},\\[1mm]
0, & \text{if } e^R(v, v') \text{ is a GPL}.
\end{cases}
\end{equation}
For each edge \((v, v')\), the associated lane set is denoted as
\begin{equation}
L(v, v') = \{ e^L(v, v'),\; e^R(v, v') \}.
\end{equation}

To enable short-horizon prediction and lane-level control, we further divide each lane into two ordered segments: an upstream segment and a downstream segment. For each edge \((v, v')\!\in\! E\), the segment set is defined as
\begin{equation}
\begin{aligned}
&S(v, v')\nonumber\\
=&
\big\{
(e^L(v, v'), 1),
(e^L(v, v'), 2), 
(e^R(v, v'), 1),
(e^R(v, v'), 2)
\big\}.
\end{aligned}
\end{equation}
At any time, each vehicle occupies exactly one segment 
\((e^\sigma(v, v'), m) \in S\), where \(\sigma \in \{L, R\}\) indicates the lane and \(m \in \{1, 2\}\) indicates the upstream or downstream segment.

We consider three types of vehicles traveling in the network: HDVs, CAVs, and buses. The corresponding sets are denoted by \(\mathcal{N}^{\text{hv}}\), \(\mathcal{N}^{\text{cav}}\), and \(\mathcal{N}^{b}\), respectively, and the overall vehicle set is
\begin{equation}
\mathcal{N} = 
\mathcal{N}^{\text{hv}} \cup 
\mathcal{N}^{\text{cav}} \cup 
\mathcal{N}^{b}.
\end{equation}

\begin{figure}
    \centering
    \includegraphics[width=1\linewidth]{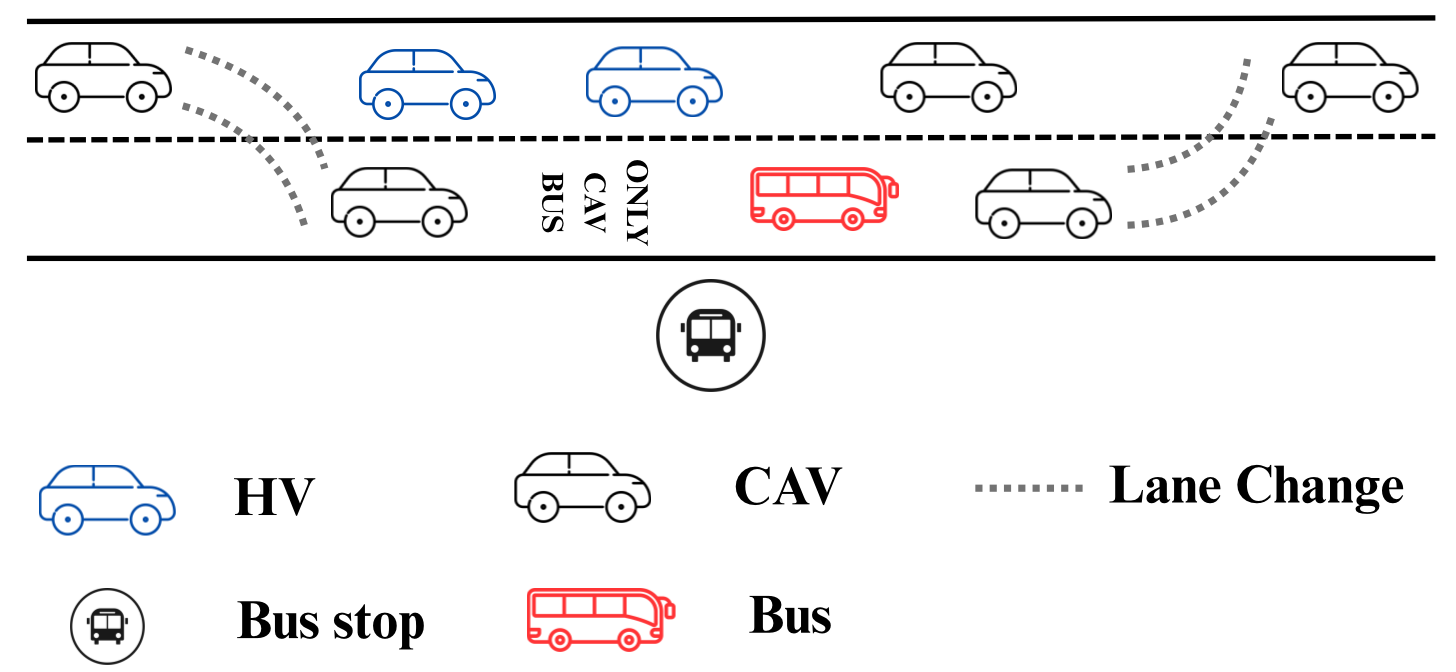}
    \caption{Illustration of the joint DL and lane-change behavior.}
    \label{fig1}
\end{figure}

As illustrated in Fig.~\ref{fig1}, buses use only DL and aim to arrive at each bus stop on time, following a specific timetable. In contrast, HDVs can use only GPLs for travel purposes and route planning. With higher flexibility, CAVs can switch between joint DLs and GPLs to improve travel efficiency while minimizing interruptions to buses.

For each bus \(b \in \mathcal{N}^{b}\), we denote its ordered sequence of visited intersections by
$
(v_{b,0}, v_{b,1}, \ldots, v_{b,N_b}),
$
where \(v_{b,0}\) and \(v_{b,N_b}\) are the origin and destination of the bus, and \(N_b \in \mathbb{N}\) denotes the number of edges on its fixed route. Since buses travel exclusively on DLs, their lane-level route is
\begin{equation}
P_b =
\big(
e^R(v_{b,0}, v_{b,1}),
\ldots,
e^R(v_{b,N_b-1}, v_{b,N_b})
\big),
\end{equation}
subject to the feasibility conditions
\begin{equation}
\delta(v_{b,k}, v_{b,k+1}) = 1,
\qquad k = 0, 1, \ldots, N_b - 1.
\end{equation}

For each CAV \(i \!\in\! \mathcal{N}^{\text{cav}}\), let \(v_{i,0}\!=\! o_i\) and \(v_{i,K_i}\!=\!d_i\) denote its origin–destination (OD) pair, where \(K_i\!\in\!\mathbb{N}\) is the number of edges on its route and
$
(v_{i,0}, v_{i,1}, \ldots, v_{i,K_i})
$
is the ordered sequence of intersections visited by CAV \(i\). Here, the index \(k\!\in \!\{0, 1, \ldots, K_i - 1\}\) labels the \(k\)-th edge \((v_{i,k}, v_{i,k+1})\) along the route of CAV \(i\). At each edge $(v_{i,k}, v_{i,k+1})$ along the route, the lane choice decision is made when CAV $i$ enters the upstream segment of this edge. At that decision-making point, CAV $i$ selects either the left or the right lane
\begin{equation}
e_i(v_{i,k}, v_{i,k+1})
\in
\left\{
e^L(v_{i,k}, v_{i,k+1}),
\;
e^R(v_{i,k}, v_{i,k+1})
\right\},\nonumber
\end{equation}
and the selected lane remains fixed while traversing the downstream segment of the same edge. This yields the path
\begin{equation}
P_{o_i, d_i}
=
\big(
e_i(v_{i,0}, v_{i,1}),
\ldots,
e_i(v_{i,K_i-1}, v_{i,K_i})
\big).
\end{equation}

Within each chosen lane \(e_i(v_{i,k}, v_{i,k+1})\), the location of CAV \(i\) is roughly represented by one of the two segments \((e_i(v_{i,k}, v_{i,k+1}), 1)\) or \((e_i(v_{i,k}, v_{i,k+1}), 2)\) based on its position, which are elements of the segment set \(S(v_{i,k}, v_{i,k+1})\). HDVs are treated as passive participants in the network, following certain routes during their trips.

In this paper, we address the coordination problem in mixed-traffic environment where CAVs dynamically utilize the DL for efficiency while buses require protected DL access to maintain schedule adherence. We propose a coordinated lane-change framework coupled with routing planning that preserves bus priority on DL segments through hard constraints, ensures path feasibility under turning and downstream compatibility requirements, and limits excessive lane-changing actions. The objective is to improve bus timeliness while maintaining efficient CAV mobility and overall traffic performance.

\section{Segment Based Modeling}
In this section, we develop a predictive, segment-resolved monitoring model for mixed traffic networks. The objective is to formally present: (i) a discrete-time prediction model that estimates inflows on each lane segment, and (ii) a bus protection mechanism that characterizes potential interference conditions on the DL.

\subsection{Segment Predictive Flow Modeling}
We consider a discrete-time system with time index $t \in \{0, \Delta t, 2\Delta t, \dots \},$ where $\Delta t > 0$ denotes the control interval. For each directed edge $(v,v')\!\in\!E$, each lane $e^\sigma(v,v')\!\in\! L(v,v')$ ($\sigma\!\in\!\{L,R\}$), and each segment index $m\!\in\!\{1,2\}$, we denote the corresponding lane segment by $(e^\sigma(v,v'),m)\in S(v,v')$. At time $t$, for each CAV $i \in \mathcal{N}^\mathrm{{cav}}$, denote its current position and speed by $x_i(t)$ and $v_i(t)$, respectively. For any target lane segment $(e^\sigma(v,v'),m)$, define $d_{i,(e^\sigma(v,v'),m)}(t)\ge 0$ as the remaining distance for vehicle $i$ to reach the entrance of that segment along its current planned path. In implementation, $d_{i,(e^\sigma(v,v'),m)}(t)$ can be computed using the current edge/lane position from the simulator and the known segment boundaries. Under a short-horizon constant speed prediction, the remaining travel time to reach the segment is denoted as
\begin{equation}
\hat{\tau}_{i,(e^\sigma(v,v'),m)}(t)
=
\frac{
d_{i,(e^\sigma(v,v'),m)}(t))
}{
v_i(t)
}.
\end{equation}

We introduce a binary indicator function that captures whether CAV $i$ is predicted to enter segment $(e^\sigma(v,v'), m)$ within the next control interval as follows:
\begin{equation}
\!\!I_{i,(e^\sigma(v,v'),m)}(t)
=
\begin{cases}
1, & \text{if } \hat{\tau}_{i,(e^\sigma(v,v'),m)}(t) \in [0, \Delta t), \\
0, & \text{otherwise}.
\end{cases}
\end{equation}
We denote as $s\!=\!(e^\sigma(v,v'),m)$ a lane segment, and all quantities indexed by $s$ inherit the corresponding definitions. We estimate the inflow rate for each DL segment by
\begin{equation}
\hat f^{\mathrm{dl}}_s(t)
=
\frac{\sum_{i\in \mathcal{N}^{\mathrm{cav}}} I_{i,s}(t)}{\Delta t}.
\end{equation}
For GPL segments, we include both CAVs and monitored HDVs in the estimated inflow via 
\begin{equation}
\hat f^{\mathrm{gpl}}_s(t)
=
\frac{\sum_{i\in \mathcal{N}^{\mathrm{cav}}} I_{i,s}(t) + \Delta \mathcal{N}^{\mathrm{hv}}_s(t)}{\Delta t},
\end{equation}
where $\Delta \mathcal{N}^{\text{hv}}_{s}(t)$ denotes the number of HDVs observed to enter segment $s$ over $[t,t+\Delta t)$ with $t$ denoting the current time. 

Let $t_0(s)$ denote the free-flow travel time of segment $s$ and let $C(s)$ denote its capacity. Given the predicted inflow, we estimate the corresponding CAV travel time using the Bureau of Public Road (BPR) function~\cite{11312449,united1964traffic}:
\begin{equation}
t^{\mathrm{dl}}_{s}(t)
=
t_{0}(s)\left(
1+\alpha\left(
\frac{\hat f^{\mathrm{dl}}_s(t)}{C(s)}
\right)^\beta
\right).
\end{equation}
For GPL segments, the travel time is estimated as
\begin{equation}\label{cavt_gpl}
t^{\mathrm{gpl}}_{s}(t)
=
t_{0}(s)\left(
1+\alpha\left(
\frac{\hat f^{\mathrm{gpl}}_s(t)}{C(s)}
\right)^\beta
\right),
\end{equation}
where $\alpha,\beta\!>\!0$ are BPR parameters. This monitoring preserves lane resolution and enables localized congestion risk detection, which is critical for joint DL operations.

\subsection{Bus Predictive Protection Modeling}
To protect bus operations on DL, we introduce a bus predictive monitoring window. For each bus $b\in \mathcal{N}^{\text{bus}}$ and each future DL segment $(e^R(v,v'),m)$ on the bus route, let $\hat{\tau}_{b,(e^R(v,v'),m)}(t)$ denote the predicted time to enter that segment from the current time $t$. A symmetric protection window centered at the predicted bus arrival is defined as
\begin{equation}
\mathcal{M}_{b,(e^R(v,v'),m)}(t)
=
\big[
\hat{\tau}_{b,(e^R(v,v'),m)}(t)
\pm \Delta T_b
\big],
\end{equation}
where $\Delta T_b\!>\!0$ denotes the bus protection horizon.

For each CAV $i \in \mathcal{N}^\mathrm{cav}$, we introduce a binary indicator that captures whether CAV $i$ is predicted to enter segment $(e^R(v,v'),m)$ within the bus protection window:
\begin{equation}
I^{\text{bus}}_{i,(e^R(v,v'),m)}(t)
=\begin{cases}1, &\begin{aligned}[t]
\text{if } \; t+\hat{\tau}_{i,(e^R(v,v'),m)}(t)\in \\
\mathcal{M}_{b,(e^R(v,v'),m)}(t),
\end{aligned}
\\
0, & \text{otherwise.}
\end{cases}
\end{equation}
Recall that, $s\!=\!(e^\sigma(v,v'),m)$ represent a lane segment, and in this setting, $\sigma\!=\!R$. The corresponding predicted bus conflict inflow rate is given by
\begin{equation}
q_s(t)
=
\frac{
\sum_{i \in \mathcal{N}^\mathrm{{cav}}}
I^{\text{bus}}_{i,s}(t)
}{
2\Delta T_b
}.
\end{equation}
Let $t_0(e^R(v,v'),m)$ denote the free-flow travel time of segment $(e^R(v,v'),m)$ and let $C(e^R(v,v'),m)$ denote its capacity.
Given the predicted conflict inflow $q_{(e^R(v,v'),m)}(t)$, we estimate the corresponding bus travel time according to the BPR function and obtain
\begin{equation}\label{eq:bpr_bus}
t_{b,s}(t)=t_{0}(s)\left(1+\alpha\left(\frac{q_s(t)}{C(s)}\right)^\beta\right).
\end{equation}
It provides a predicted strategy to detect potential bus interference on specific DL segments before it materializes.

\section{Dynamic Lane-change Approach}
In this section, we design a lane-change control strategy that proactively mitigates congestion risks while preserving bus operations. At each control time step $t$, the controller (i) enforces feasibility under bus protection constraints, (ii) selects lane changes via a segment utility, and (iii) when necessary, couples lane-change regulation with targeted rerouting.

\subsection{Bus Protection Constraints}
At each control time step $t \in \{0,\Delta t,2\Delta t,\ldots\}$, we enforce a hard bus protection constraint on the DL using the bus monitoring window. For each bus $b\in \mathcal{N}^{\text{bus}}$ and each future DL segment $s$, a bus-interference warning is triggered if
\begin{equation}\label{eq:con}
t_{b,s}(t) > (1+\lambda)\,t_0(s),
\end{equation}
where $t_{b,s}(t)$ is computed by \eqref{eq:bpr_bus}, $t_0(s)$ is the free-flow travel time, and $\lambda>0$ is an allowable tolerance.

\begin{figure}
    \centering
    \includegraphics[width=1\linewidth]{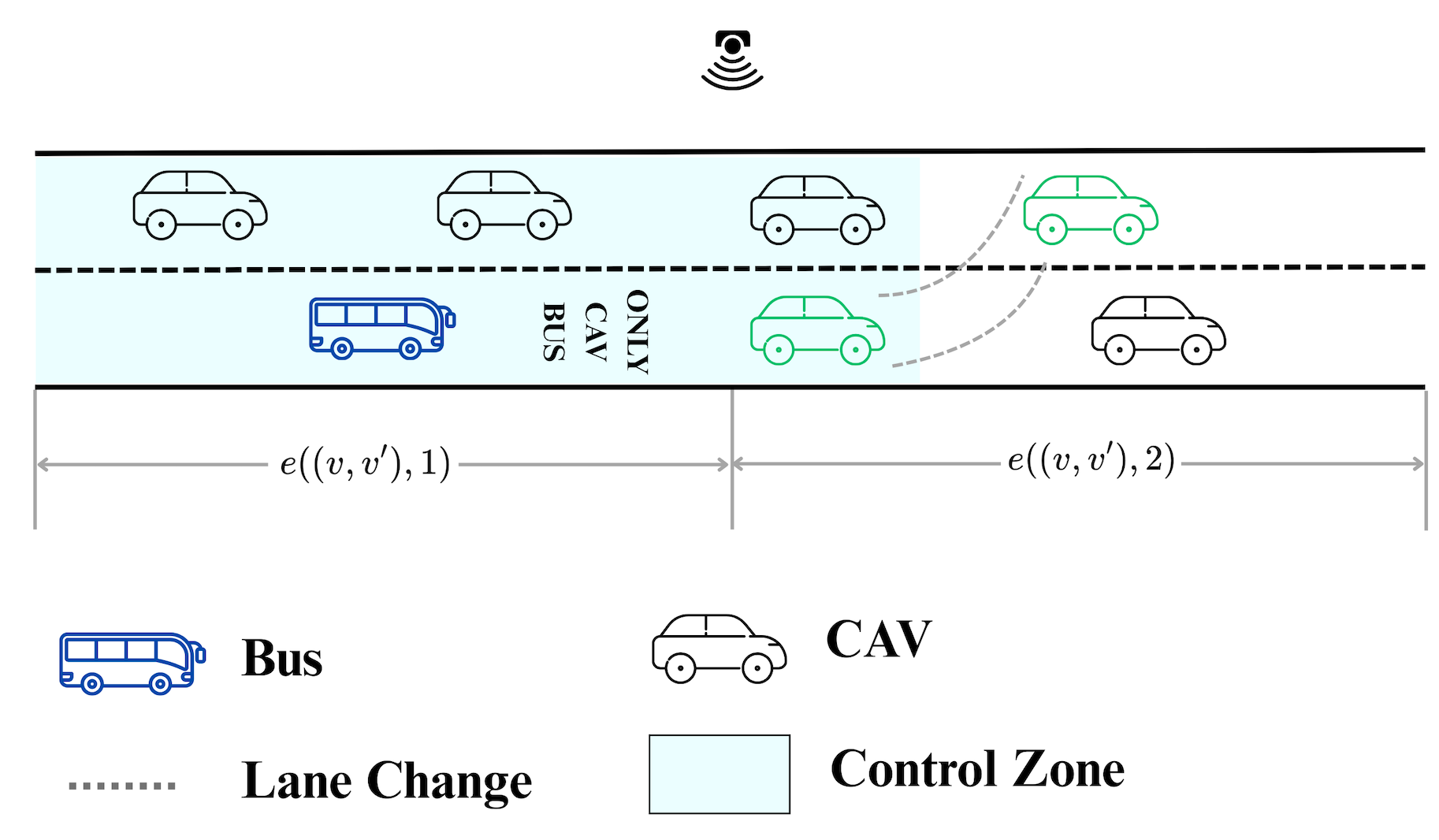}
    \caption{Segment-based monitoring and lane-change candidate selection within the control zone.}
    \label{fig2}
\end{figure}

Let $z_{i,s}(t)\in\{0,1\}$ indicate whether CAV $i$ is currently traveling on DL segment $s$ at time $t$. Let $a_i(t)\in\{-1,0,1\}$ denote the lane-change decision of CAV $i$ at time $t$, where $a_i(t)=-1$ indicates a lane change from DL to GPL, $a_i(t)=1$ indicates a lane change from GPL to DL,and $a_i(t)=0$ indicates no lane change. Whenever \eqref{eq:con} is triggered, the hard constrain enforced:
\begin{align}
\!\!\!\!a_i(t) &\!=\! -1, \quad \forall i\!\in\! \mathcal{N}^\mathrm{cav},
\ \text{s.\ t.}\ z_{i,s}(t)=1,\ I^{\text{bus}}_{i,s}(t)\!=\!1, \label{eq:exit}\\
\!\!\!\!a_i(t) &\!\neq\!1, \quad \forall i\!\in\! \mathcal{N}^\mathrm{cav},
\ \text{s.\ t.}\ z_{i,s}(t)\!=\!0,\ I^{\text{bus}}_{i,s}(t)\!=\!1. \label{eq:enter}
\end{align}
Constraint \eqref{eq:exit} forces any overlapping CAV already on the DL to move out to the GPL, as shown in Fig.~\ref{fig2}. In addition, constraint \eqref{eq:enter} prevents any overlapping CAV on the GPL from entering the DL within the same protection window.
These constraints define the admissible action set for the utility-based selection below.

\subsection{Lane-Change Control}
Lane-change decisions are evaluated independently for each lane segment $s$ at every control time step. Let $\mathcal{C}_{s}(t)\subseteq \mathcal{N}^\mathrm{cav}$ denote the candidate set on segment $s$ at time $t$. A CAV~$i$ is included in $\mathcal{C}_{s}(t)$ if and only if its action satisfies the bus protection constraints. For each candidate $i\in \mathcal{C}_{s}(t)$, we compute a weighted utility
\begin{equation}\label{eq:utility}
u_{i,s}(t)
=
w_1 \bar{u}_{1,i,s}(t)
+
w_2 \bar{u}_{2,i,s}(t)
+
w_3 \bar{u}_{3,i}(t),
\end{equation}
where $w_1,w_2,w_3\ge 0$ are tunable weights. 

Let $s'$ be the adjacent lane segment after a lane change.
Denote by $t_{s}(t)$ and $t_{s'}(t)$ the predicted traversal times of the current and adjacent segments that can be calculated. We define the following normalized travel time benefit
\begin{equation}\label{eq:u1}
\bar{u}_{1,i,s}(t)
=
\frac{t_{s}(t)-t_{s'}(t)}{t_0(s)}.
\end{equation}

\begin{figure}
    \centering
    \includegraphics[width=1\linewidth]{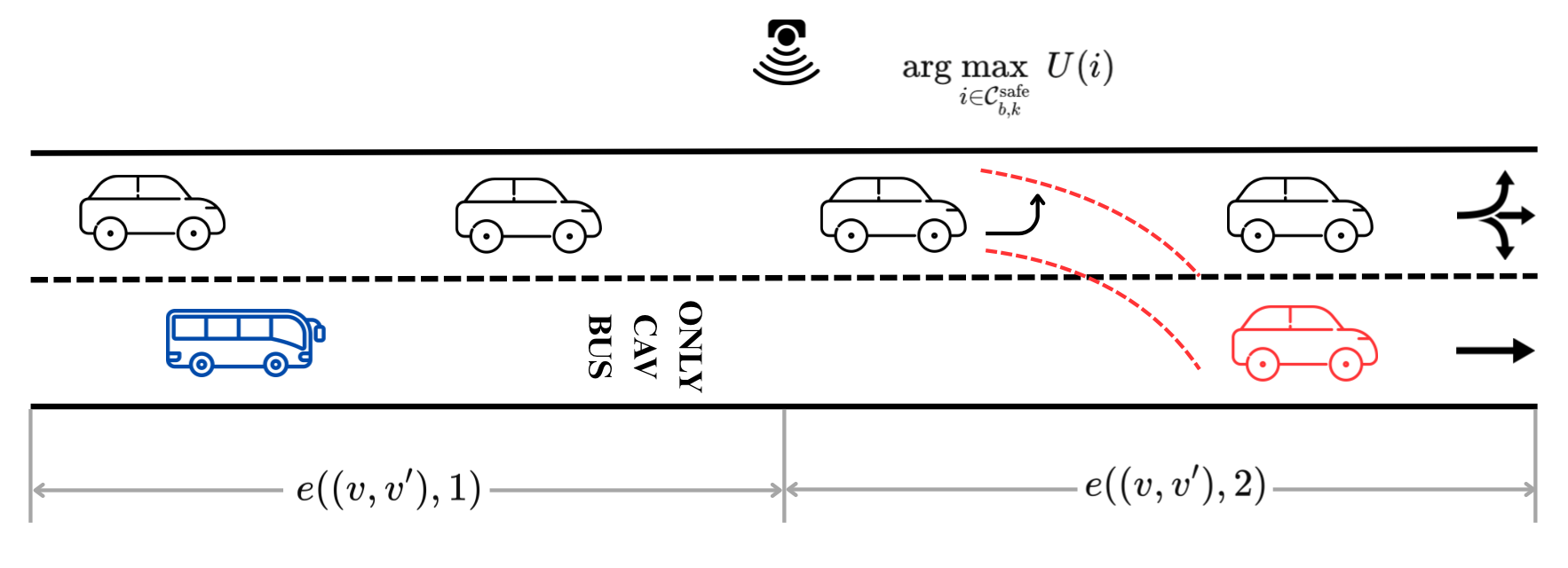}
    \caption{Predictive coupling of lane-change and routing decisions in the proposed control framework.}
    \label{fig3}
\end{figure}

Even if a lane change yields an immediate time gain, it may be undesirable if it prevents satisfying an upcoming turning requirement. Thus, we define
\begin{equation}\label{eq:u2}
\bar{u}_{2,i,s}(t)
=
\begin{cases}
1, & \text{if CAV $i$ satisfies feasibility,}\\
0, & \text{otherwise.}
\end{cases}
\end{equation}
where \emph{feasibility} means that the vehicle can still satisfy the required downstream lane positioning before reaching the next
turning point. As illustrated in Fig.~\ref{fig3}, although the CAV has sufficient space to move into the DL before the approaching bus and achieves the highest lane-change score, the behavior is prohibited due to routing inconsistency. In this example, the vehicle’s planned route requires a left-turn movement at the downstream intersection, and entering the DL would prevent it from accessing the required turning lane in time. Therefore, despite the immediate travel-time advantage, the lane change is not executed.

To discourage excessive lane-changing behavior, we penalize frequent lane changes over a rolling horizon. Let $n_i(t;T)$ denote the number of lane changes executed by CAV $i$ within the most recent $T$ seconds, over $(t-T,\,t]$. We define the lane-change rate penalty as
\begin{equation}\label{eq:u3_rate}
\bar{u}_{3,i}(t)
=
-\frac{n_i(t;T)}{T/\Delta t},
\end{equation}
which normalizes the penalty by the maximum number of control steps within a horizon of length $T$.

To avoid conflicts and limit local disturbances, at most one CAV is permitted to execute a lane change on each segment at each control step. Therefore, the controller selects
\begin{equation}\label{eq:argmax_seg}
i_s^*(t)\in \arg\max_{i\in \mathcal{C}_s(t)} u_{i,s}(t).
\end{equation}
The lane-change decision on segment $s$ at time $t$ is then given by
\begin{equation}\label{eq:lc_indicator}
N^{\mathrm{LC}}_{i,s}(t)=
\begin{cases}
1, & i=i_s^*(t)\ \text{and}\ u_{i_s^*(t),s}(t)>0,\\
0, & \text{otherwise},
\end{cases}
\end{equation}
which guarantees that the lane change is executed only by the selected vehicle and only when its score is positive.

\subsection{Rerouting Control}
Lane-change regulation alone may be insufficient when the predicted bus travel time on a protected DL segment $s$ exceeds the allowable tolerance. When the bus-interference condition is triggered, the controller evaluates the predicted travel time on the adjacent lane segment $s'$, where $s'$ denotes the neighboring GPL segment on the same edge as $s$. If the predicted travel time on $s'$ also satisfies
\begin{equation}
t_{s'}(t) > (1+\gamma)t_0(s'),
\end{equation}
where $\gamma>0$ denotes a predefined rerouting tolerance, the controller determines that lane-change regulation alone is insufficient to mitigate congestion. In this case, targeted rerouting is triggered to reduce the number of CAVs expected to reach the protected DL segment during the bus monitoring window.
We define the conflict set
\begin{equation}
\mathcal{K}_{b,s}(t)
=
\left\{ i\in \mathcal{N}^\mathrm{cav} \ \big|\ I^{\text{bus}}_{i,s}(t)=1 \right\}.
\end{equation}
A minimal subset of CAVs in $\mathcal{K}_{b,s}(t)$ is selected for rerouting such that the predicted travel time on the affected segments is reduced below the corresponding tolerance levels. Rerouting is performed by updating the remaining path of selected CAVs using a weighted shortest path search on the network, where edge costs are computed from the same short horizon prediction used for segment monitoring. This coupling mechanism enables the framework to operate at both the segment level and the network level, thereby proactively preventing DL conflicts while preserving bus schedule adherence. 

\section{Simulation}
In this section, the proposed lane-change coordination framework is implemented in a microscopic traffic simulation environment using SUMO to systematically evaluate its effectiveness in enhancing overall traffic system performance.
\subsection{Simulation Setup}
\begin{figure}
    \centering
    \includegraphics[width=1\linewidth]{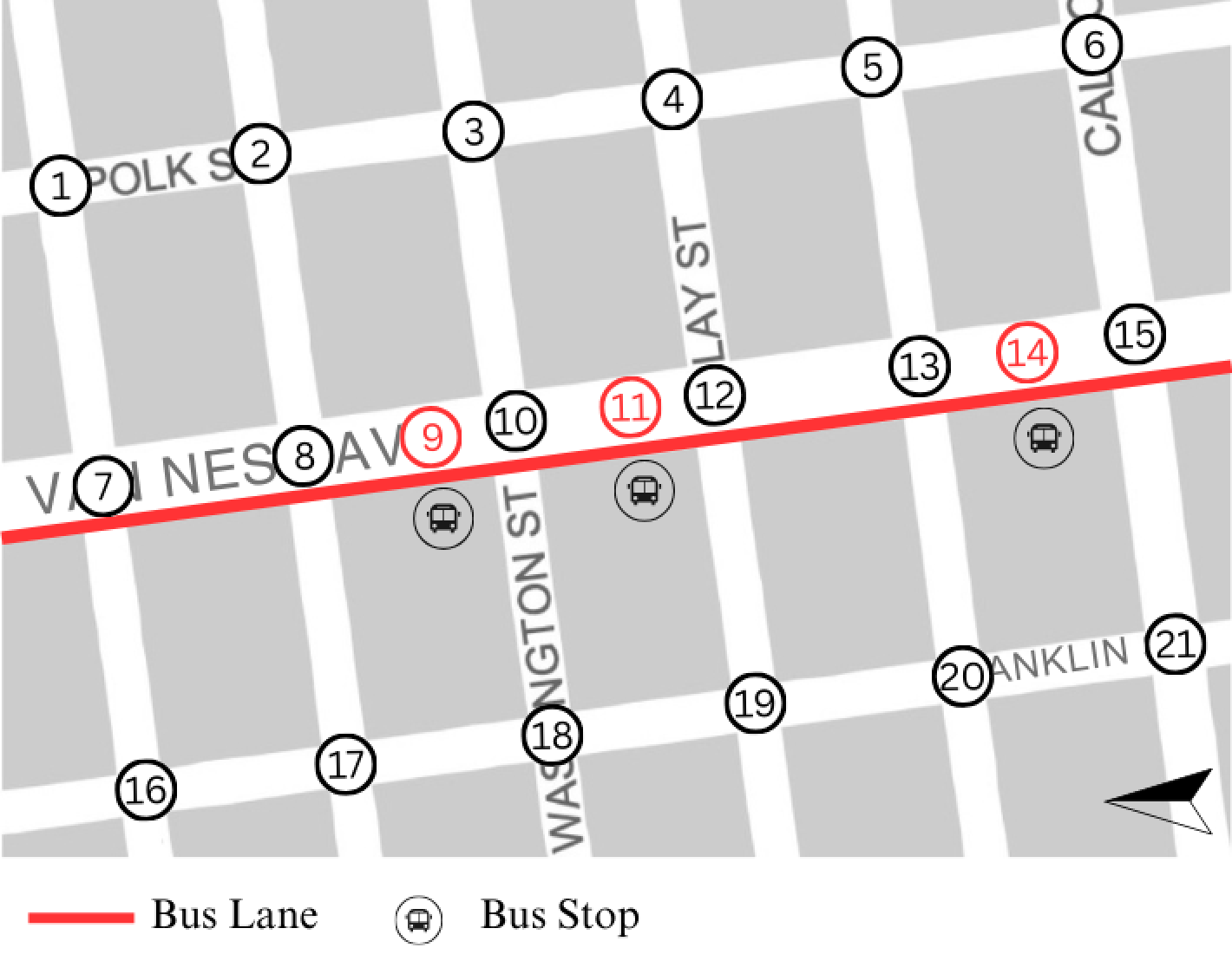}
    \caption{Study corridor on Van Ness Avenue with DL and bus stops.}
    \label{fig4}
\end{figure}

The study corridor is based on the Van Ness Avenue corridor in downtown San Francisco, spanning from Pacific Avenue to California Street and bounded by Franklin Street and Polk Street. Fig.~\ref{fig4} illustrates the real-world corridor map of the study area, highlighting the locations of bus stops, the existing DBL deployment, and its abstraction into the simulation network. The network consists of $21$ nodes, where black nodes represent signalized intersections and red nodes denote bus stop locations along the DL. The network topology and signal timing plans are extracted from OpenStreetMap. The simulation horizon is $3600~s$. During this period, $10$ buses operate along the corridor following a fixed route on the DL. Buses adhere to a predefined schedule and dwell for $60~s$ at each bus stop. CAVs and HDVs are generated from upstream boundary nodes $\{1,7,16\}$ and travel toward downstream exit nodes $\{6,15,21\}$, representing directional corridor traffic demand. A total of $700$ CAVs and $1500$ HDVs are injected into the network during the simulation horizon. CAVs adopt the intelligent driver model (IDM) with shorter desired time headways and smaller minimum gaps to capture their cooperative driving behavior. In contrast, HDVs follow the Krauss stochastic car-following model, representing conventional human driving with larger headways and greater variability \cite{krajzewicz2012sumo}.

We define the decision interval for lane-changing as $\Delta t\!=\!15$~s. In addition, a shorter prediction interval $\Delta t_b\!=\!10$~s is adopted for bus monitoring to better preserve bus priority along the DL. We compare three scenarios:
(i) dynamic routing planning (DRP): In this baseline scenario, each CAV is initially assigned to the shortest path based on real-time travel time estimation. CAVs continuously evaluate alternative routes and reroute if a shorter path is identified during traveling. A weighted Dijkstra algorithm is employed, where edge weights are initialized using free-flow travel times and dynamically updated according to real-time traffic conditions \cite{qing2017improved}. (ii) Proposed rerouting planning (PRP) proposed in our earlier work~\cite{liang2025coordinatedroutingapproachenhancing}: In this approach, proactive traffic flow is estimated within a prediction horizon, and CAVs are rerouted dynamically when a potential delay is anticipated. Unlike DRP, rerouting decisions are triggered by predicted congestion rather than instantaneous travel time comparison. (iii) Proposed method with lane-change: This approach incorporates the developed lane-change strategy in this work with PRP. Control decisions are executed at the segment level, allowing CAVs to dynamically switch between the GPL and the DL. 

\subsection{Results and Analysis}
\begin{figure}
    \centering
    \includegraphics[width=1\linewidth]{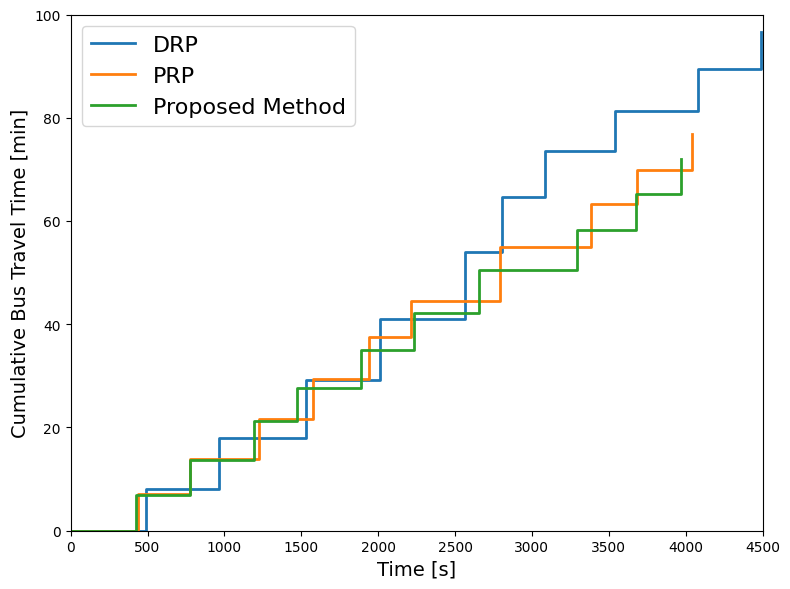}
    \caption{The cumulative bus travel time over the time horizon.}
    \label{fig5}
\end{figure}

To evaluate the effectiveness of the proposed method in mitigating bus delay, Fig.~\ref{fig5} presents the cumulative bus travel time under the three control strategies. For the proposed method, the lane-change utility weights are set as $w_1\!=\!0.3$, $w_2\!=\!0.3$, and $w_3\!=\!0.4$. It can be observed that DRP yields the highest cumulative travel time. In the early stage of the simulation, when traffic demand is relatively low, the performance of all methods is comparable. However, as congestion gradually forms, buses under DRP begin to experience increasing delays compared with PRP and the proposed method. This effect accumulates over time, resulting in the last bus under DRP completing its corridor traversal noticeably later than under the other two strategies. Both PRP and the proposed method exhibit similar performance in protecting bus operations, as both strategies proactively reserve sufficient capacity on the DL to prevent bus interference from surrounding traffic. However, due to slight differences in bus departure times, cumulative travel time curves alone do not fully reflect the actual delay mitigation performance. 

Table~\ref{tab:ontime} reports the bus delay in the three approaches. The delay is defined as the difference between the actual arrival time at each station and the scheduled arrival time, with a tolerance of $30$~s. The results show that both PRP and the proposed method nearly eliminate bus delays. Notably, the proposed method achieves a $100\%$ on-time arrival rate across all stations, demonstrating a slight improvement over PRP.

\begin{table}[t] \centering \caption{Bus on-time arrival percentage at each station.} 
\label{tab:ontime} 
\begin{tabular}{p{3.3cm}ccc} \hline Methods & Station 1 & Station 2 & Station 3 \\ \hline DPR & 40\% & 30\% & 30\% \\ PRP & 100\% & 90\% & 90\% \\ Proposed method & 100\% & 100\% & 100\% \\ \hline \end{tabular} \end{table}

\begin{figure}
    \centering
    \includegraphics[width=1\linewidth]{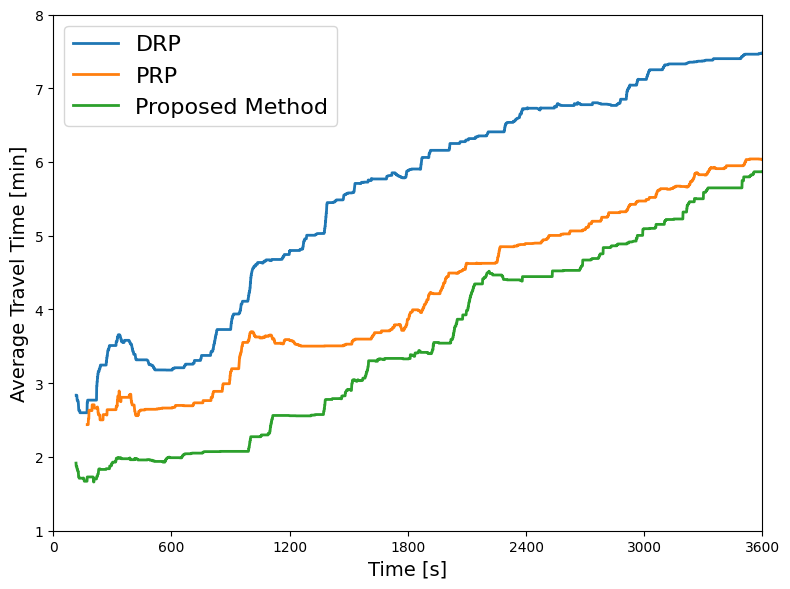}
    \caption{The average CAV travel time over time horizon.}
    \label{fig7}
\end{figure}

\begin{figure}
    \centering
    \includegraphics[width=1\linewidth]{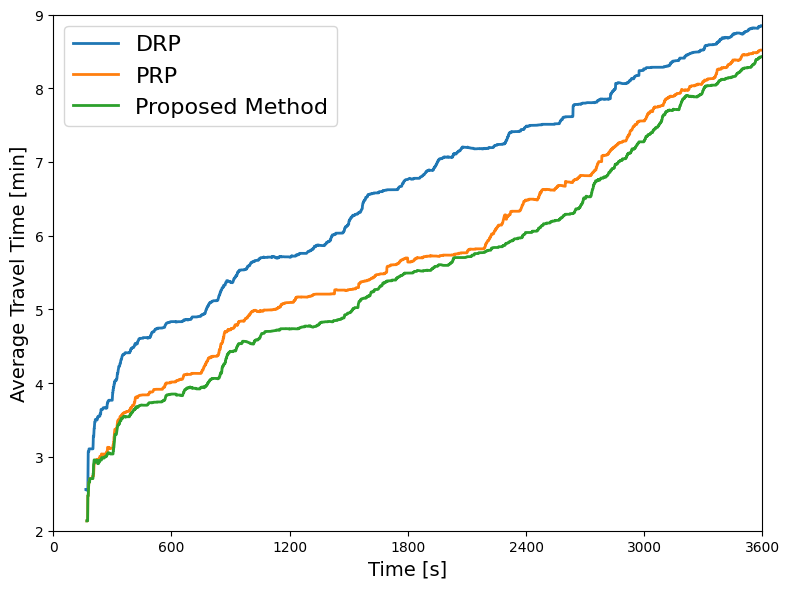}
    \caption{The average HV travel time over time horizon.}
    \label{fig8}
\end{figure}

Figs.~\ref{fig7} and~\ref{fig8} present the evolution of the average completed travel time for CAVs and HDVs, respectively. The reported average travel time is computed based on all vehicles that have completed their trips through the network up to each time instant, thereby reflecting the dynamic evolution of travel efficiency over the simulation horizon. As shown in both figures, DRP consistently results in the highest average travel time. Since DRP reacts only to instantaneous travel time variations, it cannot anticipate emerging congestion, leading to a gradual accumulation of delays. Both PRP and the proposed method significantly outperform DRP, as they proactively anticipate potential congestion through short-horizon flow prediction and adjust CAV routing before delays materialize. Although PRP and the proposed method exhibit comparable performance in protecting bus operations, the proposed method further reduces travel time for both CAVs and HDVs, with a more pronounced improvement observed for CAVs. This improvement stems from the coupling of predictive lane-change control and routing feasibility, which enables more refined utilization of the DL while avoiding unnecessary lane changes. As a result, congestion is mitigated more efficiently across the network without compromising bus priority.

\begin{figure}
    \centering
    \includegraphics[width=1\linewidth]{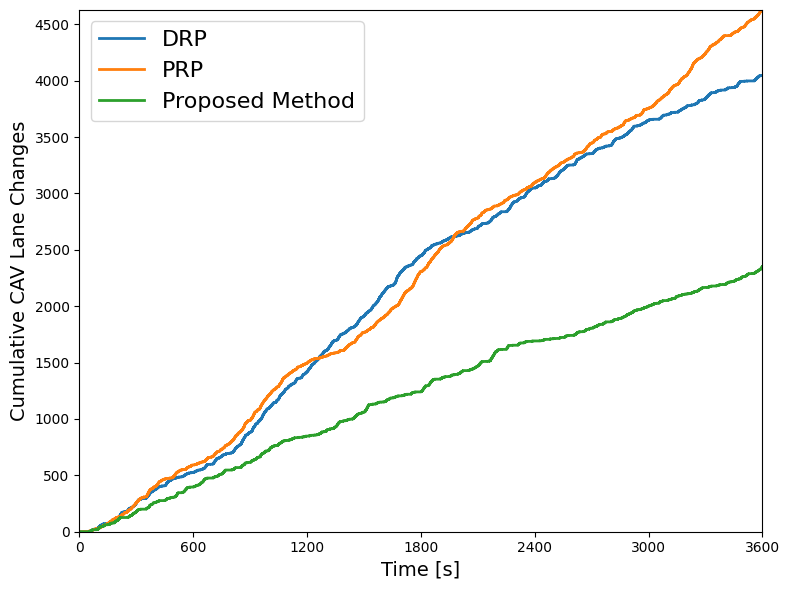}
    \caption{Cumulative CAV lane changes over the time horizon}
    \label{fig9}
\end{figure}

In both DRP and PRP, lane-change behavior is not explicitly regulated. CAVs perform lane changes in a myopic manner by selecting the adjacent lane that offers an immediate speed advantage. Such decentralized and reactive lane selection behavior may lead to frequent and unnecessary maneuvers at the network level. Fig.~\ref{fig9} compares the cumulative CAV lane changes under DRP, PRP, and the proposed method. The proposed method reduces the total number of lane changes by nearly half compared to the uncontrolled scenarios. The significant reduction in lane-changing frequency demonstrates that the predictive and score-based coordination mechanism effectively suppresses excessive or opportunistic maneuvers. Fewer lane changes not only enhance traffic stability but also reduce potential safety risks and disturbance propagation within the traffic environment.

\begin{table}[t] \centering \caption{Impact of $w_3$ on lane-change behavior} \label{tab:w3} \begin{tabular}{ccc} \toprule $w_3$ & Total CAV Lane Changes & Avg. Lane Changes per CAV \\ \midrule 0.2 & 2466 & 5.4 \\ 0.4 & 2381 & 5.1 \\ 0.6 & 2351 & 4.6 \\ \bottomrule \end{tabular} \end{table}

\begin{figure}[t]
    \centering
    \includegraphics[width=1\linewidth]{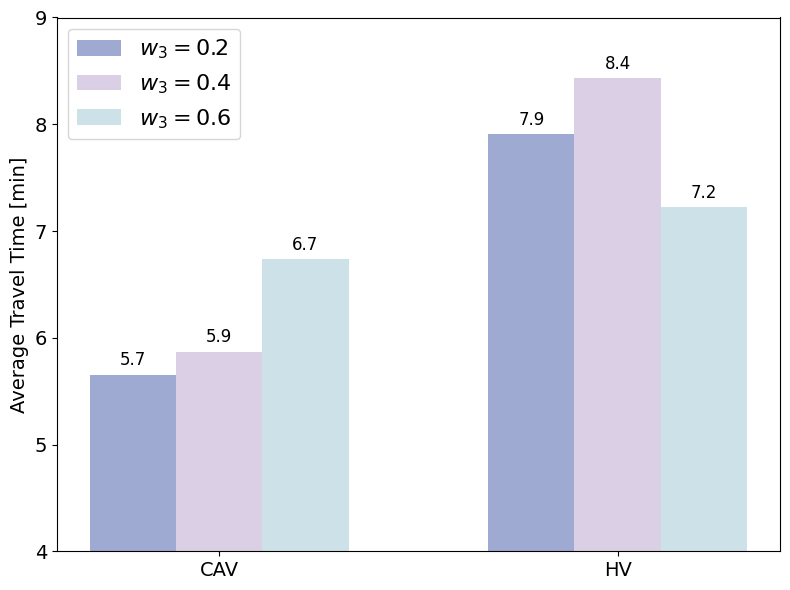}
    \caption{Impact of lane-change penalty weight $w_3$ on average travel time.}
    \label{fig10}
\end{figure}
To further evaluate the impact of the lane-change penalty weight $w_3$, which regulates the intensity of lane-change control, we test three representative values: $w_3\!=\!0.2$, $0.4$, and $0.6$. Table ~\ref{tab:w3} reports the corresponding total number of CAV lane changes and the average number of lane changes per CAV under each setting. As expected, a larger penalty weight leads to fewer lane changes, whereas a smaller weight encourages more frequent maneuvering. We continue to use average travel time as the primary efficiency metric. The impact of $w_3$ on bus performance is negligible, indicating that the bus protection mechanism remains robust across different lane-change regulation strengths. Noticeable variations are observed for CAVs and HDVs.

Fig.~\ref{fig10} shows the final average travel time of all completed vehicles. For CAVs, increasing $w_3$ leads to longer travel times, which is intuitive since stricter regulation limits their ability to opportunistically switch to faster lanes. In contrast, the impact on HDVs is non-monotonic. While $w_3\!=\!0.4$ yields the highest HV travel time, $w_3\!=\!0.6$ results in the lowest. The relationship indicates the potential existence of a critical penalty weight that balances CAV mobility and HV stability. Identifying such an operating point is essential for achieving coordinated efficiency across mixed traffic environments.

The above results validate that the proposed predictive lane-change coordination framework effectively preserves bus schedule adherence while enhancing overall network efficiency. By integrating congestion anticipation with segment-level regulation, the method mitigates potential DL conflicts before they materialize and reduces unnecessary maneuvering behavior. Compared to purely reactive or routing strategies, the coupled approach achieves a more balanced trade-off between mobility, stability, and operational reliability.


\section{Conclusions}
In this paper, we proposed a predictive lane-change control framework that coordinates CAV maneuvers in mixed traffic corridors where buses and CAVs share a DL. The approach integrates a bus protection mechanism with a weighted lane-change utility that jointly considers predicted travel time benefits, downstream routing consistency, and lane-change stability. By structuring the network at the segment level and coupling routing feasibility with lane-change decisions, the proposed strategy proactively regulates CAV behavior to preserve bus priority while maintaining overall network efficiency. Simulation studies conducted in SUMO show that the framework improves bus schedule adherence and reduces average travel time for both CAVs and HDVs. Future work will focus on identifying optimal operating points on each edge within the proposed framework and evaluating the control performance of the method in large-scale urban networks.


\bibliographystyle{IEEEtran}
\bibliography{ITSC_2025,IDS}
\end{document}